\documentclass[prd,showpacs,showkeys,nofootinbib,twocolumn]{revtex4-1}

\usepackage{graphicx}
\usepackage{amsmath,amsfonts,amssymb}
\usepackage[utf8]{inputenc}

\DeclareUnicodeCharacter{2212}{-}

\begin{document}

\title{Gravitational clock compass and the detection of gravitational waves}

\author{Peter A. Hogan}
\email{peter.hogan@ucd.ie}
\affiliation{School of Physics, University College Dublin, Belfield, Dublin 4, Ireland} 

\author{Dirk Puetzfeld}
\email{dirk.puetzfeld@zarm.uni-bremen.de}
\homepage{http://puetzfeld.org}
\affiliation{University of Bremen, Center of Applied Space Technology and Microgravity (ZARM), 28359 Bremen, Germany} 

\date{ \today}

\begin{abstract}
We present an alternative derivation of the gravitational clock compass and show how such a device can be used for the detection of gravitational waves. Explicit compass setups are constructed in special types of space--times, namely for exact plane gravitational waves and for waves moving radially relative to an observer.  
\end{abstract}


\maketitle


\section{Introduction}\label{introduction_sec}

Modern clocks reached an unprecedented level of accuracy and stability \cite{Chou:etal:2010,Huntemann:etal:2012,Guena:etal:2012,Falke:etal:2014,Bloom:etal:2013,Schioppo:etal:2016,Bauch:2019} in recent years. Therefore it appears obvious to utilize them for a direct detection of the gravitational field.

Building upon a preceding series of works \cite{Puetzfeld:Obukhov:2016:1,Puetzfeld:Obukhov:2018:1,Puetzfeld:Obukhov:2018:2} -- in which we derived general prescriptions for the setup of the constituents of a device called a ``gravitational compass'' \cite{Szekeres:1965} and a ``clock compass'', i.e.\ realizations of gradiometers in the context of the theory of General Relativity -- we are now going to study how clocks can be used in an operational way to explicitly map gravitational wave space--times by means of mutual frequency comparisons.

In a previous work \cite{Puetzfeld:Obukhov:2016:1} on the standard gravitational compass we employed a covariant expansion technique based on Synge's world function \cite{Synge:1960,DeWitt:Brehme:1960}, while in the context of the clock compass \cite{Puetzfeld:Obukhov:2018:1} the derivation was based on the construction of a suitable normal coordinate system. Here we present an alternative approximation technique, motivated by earlier work on radiation from isolated systems \cite{Hogan:Trautman:1987} and on work on the equations of motion in General Relativity \cite{Hogan:Futamase:Itoh:2008,Hogan:Asada:Futamase:2010}. It offers a different perspective on the derivation of the measurable frequency ratio between the clocks and is not, like \cite{Puetzfeld:Obukhov:2018:1}, based on \cite{Hehl:Ni:1990} as a starting point.  

Implementing the gravitational compass in the cases of explicit gravitational field models involves working in a coordinate system since the models are expressed in a coordinate system. This paper demonstrates how one might do this since it is in a coordinate system which (i) respects, or is tailored to, the geometry and (ii) applies to any model since it involves six functions of the four coordinates. The examples we describe demonstrate that even working in a suitable coordinate system the application to explicit models involving gravitational radiation is a complex procedure.

The structure of the paper is as follows: In section \ref{sec_minkowski} we review a suitable form of the flat space--time metric around a world line along in a reference frame carried by a general observer. In the subsequent sections \ref{sec_freq_ratio_flat} and \ref{sec_freq_ratio_curved} an expression for the frequency ratio between a clock carried by the observer and a clock in the vicinity of his world line is given in a flat as well as in a curved background. This is followed by a derivation of the frequency ratio in a plane gravitational wave background in the sections \ref{sec_plane_waves_1} and \ref{sec_plane_waves_2}, as well as for radial waves in section \ref{sec_radial_waves}. In section \ref{sec_clock_compass} we show how an ensemble of clocks has to be prepared in order to allow for a measurement of all independent components of the curvature tensor through mutual frequency comparisons of the clocks in the previously derived space--times. We draw our conclusions in section \ref{sec_conclusions}. Appendix \ref{sec_notation} contains a brief overview of the notations and conventions used throughout the article, whereas some consistency checks are given in appendix \ref{sec_consistency_curvature}. 

\section{Minkowskian Preliminaries}\label{sec_minkowski}

In this section we demonstrate how to construct a coordinate system for Minkowskian space-time based on a family of space-like hypersurfaces since the extension of this construction to general curved space-times is the central feature of this paper. The construction in this section leads to a particularly useful form (eq.\ (\ref{25})) of the Minkowskian line element which is important for the derivation of the frequency ratio in flat space-time in the next section. In addition basic formulas (eqs.\ (\ref{51})-(\ref{55})) in the Minkowskian case play an important role in the neighbourhood of a time-like world line in the general curved space-time case later.

\begin{figure}
\begin{center}
\includegraphics[width=7cm,angle=-90]{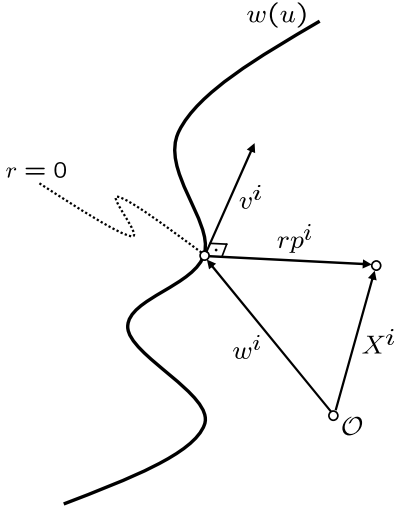}
\end{center}
\caption{\label{fig_0} Construction of the coordinates of a point $X$ in the vicinity the time-like world line $w(u)$ parametrized by the proper time $u$. The parameter $r$ is centered on the world line, and the space-like vector $p^i$ is chosen to be orthogonal to the velocity $v^i$ along the world line.}
\end{figure}

We begin with the Minkowskian line element in rectangular Cartesian coordinates and time $X^i=(X, Y, Z, T)$:
\begin{equation}\label{1}
ds^2=-(dX)^2-(dY)^2-(dZ)^2+(dT)^2=\eta_{ij}\,dX^i\,dX^j\ .
\end{equation} 
Writing
\begin{equation}\label{2}
X^i=w^i(u)+r\,p^i(u)\ ,
\end{equation}
with 
\begin{equation}\label{3}
p_i\,p^i=-1\ ,
\end{equation}
we see that $p^i$ is a unit space--like vector field defined along the world line, which we take to be time--like with $u$ taken to be proper--time or arc length along it. Hence 
\begin{equation}\label{4}
v^i(u)=\frac{dw^i}{du}\ \ \ {\rm with}\ \ \ v_i\,v^i=+1\ .
\end{equation}
Thus $v^i(u)$ is the unit time--like tangent vector field to $r=0$ and is therefore the 4--velocity of an observer with world line $r=0$, see figure \ref{fig_0} for a sketch of the setup. The 4--acceleration of an observer with world line $r=0$ is 
\begin{equation}\label{5}
a^i(u)=\frac{dv^i}{du}\ ,
\end{equation}
and this satisfies $a_i\,v^i=0$ on account of the second of (\ref{4}). The unit space--like vector field $p^i$ defined along $r=0$ is assumed to be orthogonal to $r=0$ at each of its points and thus
\begin{equation}\label{6}
v_i\,p^i=0\ .
\end{equation}
We are free to choose the transport law for $p^i$ along $r=0$ subject to ensuring that (\ref{3}) and (\ref{6}) are preserved at all points of $r=0$. For our present purposes we construct the transport law for $p^i$ as follows: Begin by defining an orthonormal tetrad $\{\lambda^i_{(a)}\}$ with $a=1, 2, 3, 4$, at a point of $r=0$ with
\begin{equation}\label{7}
\eta_{ij}\,\lambda^i_{(a)}\,\lambda^j_{(b)}=\eta_{(a)(b)}={\rm diag}(-1, -1, -1, +1)\ .
\end{equation}
Tetrad indices (or labels) will be those indices with round brackets around them. They will be raised and lowered with $\eta^{(a)(b)}$ and $\eta_{(a)(b)}$ respectively with the former defined by $\eta_{(a)(b)}\,\eta^{(b)(c)}=\delta^c_a$. We shall choose
\begin{equation}\label{8}
\lambda^i_{(4)}=v^i\ ,
\end{equation}
and we can invert (\ref{7}) to read
\begin{equation}\label{9}
\eta^{ij}=\eta^{(a)(b)}\,\lambda^i_{(a)}\,\lambda^j_{(b)}=-\lambda^i_{(\alpha)}\,\lambda^j_{(\alpha)}+v^i\,v^j\ ,
\end{equation}
with henceforth Greek indices taking values 1, 2, 3. On account of (\ref{8}) $\lambda^i_{(4)}$ is extended to a field on $r=0$ since we shall take (\ref{8}) to hold at all points of $r=0$ and thus 
\begin{equation}\label{10}
\lambda^i_{(4)}(u)=v^i(u)\ ,
\end{equation}
for all $u$. We define $\lambda^i_{(\alpha)}(u)$ along $r=0$ by requiring this orthonormal triad to be transported according to the law:
\begin{equation}\label{11}
\frac{d\lambda^i_{(\alpha)}}{du}=-v^i\,a^j\,\lambda_{(\alpha)j}+\omega^{ij}\,\lambda_{(\alpha)j}\ ,
\end{equation}
for $\alpha=1, 2, 3$. Here $\omega^{ij}(u)=-\omega^{ji}(u)$ and $\omega^{ij}\,v_j=0$. The first term on the right hand side of (\ref{11}) represents Fermi--Walker transport while the second term represents transport with rigid rotation. The transport law (\ref{11}) preserves the scalar products
\begin{equation}\label{12}
\eta_{ij}\,\lambda^i_{(\alpha)}\,\lambda^j_{(\beta)}=-\delta_{\alpha \beta}\ \ \ {\rm and}\ \ \ v_i\,\lambda^i_{(\alpha)}=0\ ,
\end{equation}
along $r=0$. Multiplying (\ref{9}) by $p_j$ we have, on account of (\ref{6}),
\begin{equation}\label{13}
p^i=-p_{(\alpha)}\,\lambda^i_{(\alpha)}=p^{(\alpha)}\,\lambda^i_{(\alpha)}\ ,
\end{equation}
and multiply (\ref{9}) by $a_j$ results in 
\begin{equation}\label{14}
a^i=-a_{(\alpha)}\,\lambda^i_{(\alpha)}=a^{(\alpha)}\,\lambda^i_{(\alpha)}\ ,
\end{equation}
with $p_{(\alpha)}=p_i\,\lambda^i_{(\alpha)}$and $a_{(\alpha)}=a_i\,\lambda^i_{(\alpha)}$. For future reference we note that
\begin{equation}\label{15}
\delta_{\alpha \beta} \, p^{(\alpha)}\,p^{(\beta)}=+1\ \ \ {\rm and}\ \ \ a^i\,p_i=-a_{(\alpha)}\,p_{(\alpha)}=a_{(\alpha)}\,p^{(\alpha)}.
\end{equation}
We will now take $p^{(\alpha)}$ to be independent of the proper time $u$ and parametrize $p^{(\alpha)}$ with the stereographic variables $x, y$ (with $-\infty<x, y<+\infty$) as
\begin{equation}\label{16}
p^{(1)}=P_0^{-1}x\ ,\ p^{(2)}=P_0^{-1}y\ ,\ p^{(3)}=P_0^{-1}\left (\frac{1}{4}(x^2+y^2)-1\right )\ ,
\end{equation}
with
\begin{equation}\label{17}
P_0=1+\frac{1}{4}(x^2+y^2)\ .
\end{equation}
It now follows from (\ref{11}), (\ref{13}) and (\ref{16}) that $p^i$ obeys, along $r=0$, the transport law
\begin{equation}\label{18}
\frac{\partial p^i}{\partial u}=-v^i\,(a^j\,p_j)+\omega^{ij}\,p_j\ .
\end{equation}
Here again the first term on the right hand side is Fermi--Walker transport while the second term represents a rigid rotation.

We now have
\begin{eqnarray}
\frac{\partial p^i}{\partial x}\,\frac{\partial p_i}{\partial x}&=&-\delta_{\alpha \beta} \frac{\partial p^{(\alpha)}}{\partial x}\,\frac{\partial p^{(\beta)}}{\partial x}=-P_0^{-2}\ , \label{19}\\
\frac{\partial p^i}{\partial y}\,\frac{\partial p_i}{\partial y}&=&-\delta_{\alpha \beta} \frac{\partial p^{(\alpha)}}{\partial y}\,\frac{\partial p^{(\beta)}}{\partial y}=-P_0^{-2}\ , \label{20_0} \\
\frac{\partial p^i}{\partial x}\,\frac{\partial p_i}{\partial y}&=&-\delta_{\alpha\beta}\frac{\partial p^{(\alpha)}}{\partial x}\,\frac{\partial p^{(\beta)}}{\partial y}=0\ .\label{20} 
\end{eqnarray}
Using (\ref{18}) we find that 
\begin{eqnarray}
\frac{\partial p_i}{\partial x}\,\frac{\partial p^i}{\partial u}&=&\omega_{(\alpha)(\beta)}\,\frac{\partial p^{(\alpha)}}{\partial x}\,p^{(\beta)}\ , \nonumber \\
\frac{\partial p_i}{\partial y}\,\frac{\partial p^i}{\partial u}&=&\omega_{(\alpha)(\beta)}\,\frac{\partial p^{(\alpha)}}{\partial y}\,p^{(\beta)}\ .\label{20a} 
\end{eqnarray}
It will also be useful to have the formulae:
\begin{equation}\label{21}
P_0^2\left (\frac{\partial p^{(\alpha)}}{\partial x}\frac{\partial p^{(\beta)}}{\partial x}+\frac{\partial p^{(\alpha)}}{\partial y}\frac{\partial p^{(\beta)}}{\partial y}\right )=\delta^{\alpha\beta}-p^{(\alpha)}\,p^{(\beta)}\ ,\end{equation}
and
\begin{eqnarray}
\frac{\partial^2p^{(\alpha)}}{\partial x^2}&=&-P_0^{-2}p^{(\alpha)}-P_0^{-1}\frac{\partial P_0}{\partial x}\,\frac{\partial p^{(\alpha)}}{\partial x}+P_0^{-1}\frac{\partial P_0}{\partial y}\,\frac{\partial p^{(\alpha)}}{\partial y}\ , \nonumber\\
\label{21a}\\
\frac{\partial^2p^{(\alpha)}}{\partial y^2}&=&-P_0^{-2}p^{(\alpha)}+P_0^{-1}\frac{\partial P_0}{\partial x}\,\frac{\partial p^{(\alpha)}}{\partial x}-P_0^{-1}\frac{\partial P_0}{\partial y}\,\frac{\partial p^{(\alpha)}}{\partial y}\ , \nonumber\\ 
\label{21b}\\
\frac{\partial^2p^{(\alpha)}}{\partial x\partial y}&=&-P_0^{-1}\frac{\partial P_0}{\partial y}\,\frac{\partial p^{(\alpha)}}{\partial x}-P_0^{-1}\frac{\partial P_0}{\partial x}\,\frac{\partial p^{(\alpha)}}{\partial y}\ ,\label{21c}
\end{eqnarray}
which can be verified by direct substitution from (\ref{16}) or otherwise.

Now (\ref{2}) gives $X^i$ in terms of $x, y, r, u$. To obtain the Minkowskian line element in coordinates $x, y, r, u$ we first have from (\ref{2})
\begin{equation}\label{22}
dX^i=\left (v^i+r\,\frac{\partial p^i}{\partial u}\right )\,du+p^i\,dr+r\,\frac{\partial p^i}{\partial x}\,dx+r\,\frac{\partial p^i}{\partial y}\,dy\ .
\end{equation}
Now using the scalar products (\ref{19})--(\ref{20a}) the Minkowskian line element (\ref{1}) becomes
\begin{eqnarray}
ds^2&=&\eta_{ij}\,dX^i\,dX^j\ ,\nonumber\\
&=&-r^2P_0^{-2}(dx^2+dy^2)+2\,r^2\omega_{(\alpha)(\beta)}\,\frac{\partial p^{(\alpha)}}{\partial x}\,p^{(\beta)}\,du\,dx\nonumber\\
&&+2\,r^2\omega_{(\alpha)(\beta)}\,\frac{\partial p^{(\alpha)}}{\partial y}\,p^{(\beta)}\,du\,dy-dr^2\nonumber\\
&&+\{(1-h_0\,r)^2-r^2\omega_{(\sigma)(\alpha)}\,\omega_{(\sigma)(\beta)}\,p^{(\alpha)}\,p^{(\beta)}\}du^2\ ,\label{23}
\end{eqnarray}
with
\begin{equation}\label{24}
h_0=a_i\,p^i=a_{(\alpha)}\,p^{(\alpha)}\ ,
\end{equation}
by (\ref{15}). We can rewrite (\ref{23}) in the neat form
\begin{eqnarray}
ds^2&=&-r^2P_0^{-2}\{(dx+a_0\,du)^2+(dy+b_0\,du)^2\} \nonumber\\
&&-dr^2+(1-h_0r)^2\,du^2\ ,\label{25}
\end{eqnarray}
with 
\begin{eqnarray}
a_0&=&-P_0^2\omega_{(\alpha)(\beta)}\,\frac{\partial p^{(\alpha)}}{\partial x}\,p^{(\beta)}\ , \nonumber \\ 
b_0&=&-P_0^2\omega_{(\alpha)(\beta)}\,\frac{\partial p^{(\alpha)}}{\partial y}\,p^{(\beta)}\ , \label{26}
\end{eqnarray}
since
\begin{eqnarray}
&&P_0^2\left (\omega_{(\alpha)(\beta)}\,\frac{\partial p^{(\alpha)}}{\partial x}\,p^{(\beta)}\right )^2+P_0^2\left (\omega_{(\alpha)(\beta)}\,\frac{\partial p^{(\alpha)}}{\partial x}\,p^{(\beta)}\right )^2 \nonumber \\
&&=\delta^{\sigma\rho}\omega_{(\sigma)(\alpha)}\,\omega_{(\rho)(\beta)}\,p^{(\alpha)}\,p^{(\beta)}\ .\label{27}
\end{eqnarray}
The last equation here is a consequence of (\ref{21}). Using (\ref{21a}) and (\ref{21b}) we see that 
\begin{equation}\label{28}
\frac{\partial a_0}{\partial x}=\frac{\partial b_0}{\partial y}\ ,
\end{equation}
and so it follows that we can write
\begin{equation}\label{29}
a_0=\frac{\partial q}{\partial y}\ \ {\rm and}\ \ b_0=\frac{\partial q}{\partial x}\ ,
\end{equation} 
with the function $q$ given by
\begin{eqnarray}
&&q(x, y, u)=\frac{1}{2}(x^2-y^2)\,\omega_{(1)(2)}+y\left (1+\frac{1}{4}x^2 -\frac{1}{12}y^2\right)\nonumber\\
&&\omega_{(1)(3)}+x\left (1-\frac{1}{12}x^2+\frac{1}{4}y^2\right )\,\omega_{(2)(3)}.\label{30}
\end{eqnarray}

For future convenience we define the 3--vectors
\begin{eqnarray}
\vec p&=&(p^{(1)}, p^{(2)}, p^{(3)})\ , \quad \vec a=(a^{(1)}, a^{(2)}, a^{(3)})\ ,\nonumber \\
 \nonumber \\
\vec\omega&=&(\omega^{(2)(3)}, \omega^{(3)(1)}, \omega^{(1)(2)})\ .\label{31}
\end{eqnarray}
Using the standard notation of the scalar product (or ``dot product") of 3--vectors and for the vector product (or ``cross product") of 3--vectors we have
\begin{eqnarray}
h_0&=&-\vec a\cdot\vec p\ ,\ \ a_0=-P_0^2\frac{\partial p^{(\alpha)}}{\partial x}\,(\vec\omega\times\vec p)^{(\alpha)}\ ,\nonumber \\
 b_0&=&-P_0^2\frac{\partial p^{(\alpha)}}{\partial y}\,(\vec\omega\times\vec p)^{(\alpha)}\ ,\label{32}
\end{eqnarray}
from which we find, using (\ref{21}), that
\begin{equation}\label{32_1}
P_0^{-2}(a_0^2+b_0^2)=(\vec\omega\times\vec p)\cdot(\vec\omega\times\vec p)=|\vec\omega\times\vec p|^2\ .
\end{equation}

In the light of the foregoing we can now say that (\ref{2}), written more explicitly, reads
\begin{equation}\label{33}
X^i=w^i(u)+r\,p^i(x, y, u)\ ,
\end{equation}
which implicitly determines $x, y, r, u$ as scalar functions of $X^i$ on Minkowskian space--time. We will need the gradients of these functions with respect to $X^i$, denoted by a comma in each case. To obtain these we start by differentiating (\ref{33}) with respect to $X^j$ giving
\begin{equation}\label{34}
\delta^i_j=\left (v^i+r\,\frac{\partial p^i}{\partial u}\right )u_{,j}+p^i\,r_{,j}+r\,\frac{\partial p^i}{\partial x}\,x_{,j}+r\,\frac{\partial p^i}{\partial y}\,y_{,j}\ .
\end{equation}
Multiplying this by $v_i, p_i, \partial p_i/\partial x$ and $\partial p_i/\partial y$ yields successively
\begin{eqnarray}
v_j&=&(1-h_0\,r)u_{,j}\ \ \Rightarrow\ \ u_{,j}=(1-h_0\,r)^{-1}v_j\ ,\label{35} \\
p_j&=&-r_{,j}\ \ \ \Rightarrow\ \ \ r_{,j}=-p_j\ ,\label{36}\\
\frac{\partial p_j}{\partial x}&=&r\,\frac{\partial p_i}{\partial x}\,\frac{\partial p^i}{\partial u}\,u_{,j}-r\,P_0^{-2}x_{,j}\ \nonumber\\
 &\Rightarrow&\ \ x_{,j}+a_0\,u_{,j}=-\frac{P_0^2}{r}\,\frac{\partial p_j}{\partial x}\ , \label{37} \\
\frac{\partial p_j}{\partial y}&=&r\,\frac{\partial p_i}{\partial y}\,\frac{\partial p^i}{\partial u}\,u_{,j}-r\,P_0^{-2}y_{,j}\ \nonumber \\
&\Rightarrow&\ \ y_{,j}+b_0\,u_{,j}=-\frac{P_0^2}{r}\,\frac{\partial p_j}{\partial y}\ ,\label{38} 
\end{eqnarray}
with the final two cases relying on (\ref{20a}) and (\ref{26}). We first note from (\ref{35})--(\ref{38}) that
\begin{equation}\label{39}
p^i\,\frac{\partial}{\partial X^i}=p^i\left (x_{,i}\frac{\partial}{\partial x}+y_{,i}\frac{\partial}{\partial y}+r_{,i}\frac{\partial}{\partial r}+u_{,i}\frac{\partial}{\partial u}\right )=\frac{\partial}{\partial r}\ .
\end{equation}
Substituting (\ref{35})--(\ref{38}) back into (\ref{34}), using (\ref{18}) and raising the covariant index using $\eta^{ij}$, we have
\begin{eqnarray}
&&\eta^{ij}=-P_0^{2}\left (\frac{\partial p^i}{\partial x}\frac{\partial p^j}{\partial x}+\frac{\partial p^i}{\partial y}\frac{\partial p^j}{\partial y}\right )-p^i\,p^j+v^i\,v^j\nonumber\\
&&+r(1-h_0\,r)^{-1}\left\{\omega^{ik}\,p_k-a_0\,\frac{\partial p^i}{\partial x}-b_0\,\frac{\partial p^i}{\partial y}\right\}v^j\ .\label{40}
\end{eqnarray}
However the final term here vanishes since
\begin{eqnarray}
&&\omega^{ik}\,p_k-a_0\,\frac{\partial p^i}{\partial x}-b_0\,\frac{\partial p^i}{\partial y} \nonumber \\
&&=\left (\delta^{\beta \gamma} \omega_{(\alpha)(\beta)}\,p_{(\gamma)}+a_0\,\frac{\partial p_{(\alpha)}}{\partial x}+
b_0\,\frac{\partial p_{(\alpha)}}{\partial y}\right )\lambda^{i(\alpha)},\label{41}
\end{eqnarray}
and
\begin{eqnarray}
&&a_0\,\frac{\partial p^{(\alpha)}}{\partial x}+b_0\,\frac{\partial p^{(\alpha)}}{\partial y} \nonumber\\
&=&P_0^2\omega_{(\gamma)(\beta)}\,p^{(\beta)}\left (\frac{\partial p^{(\gamma)}}{\partial x}
\frac{\partial p^{(\alpha)}}{\partial x}+\frac{\partial p^{(\gamma)}}{\partial y}\frac{\partial p^{(\alpha)}}{\partial y}\right )\nonumber\\
&=&\omega_{(\gamma)(\beta)}\,p^{(\beta)}(\delta^{\gamma\alpha}-p^{(\gamma)}\,p^{(\alpha)}) \nonumber\\
&\stackrel{(\ref{21})}{=}&-\omega^{(\alpha)(\beta)}\,p_{(\beta)} .\label{42}
\end{eqnarray}
Hence (\ref{40}) simplifies to 
\begin{equation}\label{43}
\eta^{ij}=-P_0^{2}\left (\frac{\partial p^i}{\partial x}\frac{\partial p^j}{\partial x}+\frac{\partial p^i}{\partial y}\frac{\partial p^j}{\partial y}\right )-p^i\,p^j+v^i\,v^j\ .
\end{equation}

The line element (\ref{25}) can be written
\begin{equation}\label{44}
ds^2=-(\vartheta^{(1)})^2-(\vartheta^{(2)})^2-(\vartheta^{(3)})^2+(\vartheta^{(4)})^2\ ,
\end{equation}
with the basis 1--forms given by
\begin{eqnarray}
\vartheta^{(1)}&=&r\,P_0^{-1}(dx+a_0\,du)=r\,P_0^{-1}(x_{,i}+a_0\,u_{,i})dX^i\nonumber\\
&=&-P_0\,\frac{\partial p_i}{\partial x}\,dX^i=-P_0\frac{\partial p^{(\alpha)}}{\partial x}\,\lambda_{(\alpha)i}\,dX^i\, ,\label{45}\\
\vartheta^{(2)}&=&r\,P_0^{-1}(dy+b_0\,du)=r\,P_0^{-1}(y_{,i}+b_0\,u_{,i})dX^i\nonumber\\
&=&-P_0\,\frac{\partial p_i}{\partial y}\,dX^i=-P_0\,\frac{\partial p^{(\alpha)}}{\partial y}\,\lambda_{(\alpha)i}\,dX^i\ ,\label{46}\\
\vartheta^{(3)}&=&dr=r_{,i}\,dX^i=-p_i\,dX^i\nonumber\\
&=&-p^{(\alpha)}\,\lambda_{(\alpha)i}\,dX^i\ ,\label{47}\\
\vartheta^{(4)}&=&(1-r\,h_0)\,du=(1-r\,h_0)\,u_{,i}\,dX^i\nonumber\\
&=&v_i\,dX^i\ .\label{48}
\end{eqnarray}

Finally we shall also require $p_{i,j}$ and we can obtain this by rewriting (\ref{34}), using (\ref{35}) and (\ref{36}), as
\begin{equation}\label{49}
\delta^i_j=v^i\,u_{,j}+p^i\,r_{,j}+r\,p^i{}_{,j}=(1-r\,h_0)^{-1}v^i\,v_j-p^i\,p_j+r\,p^i{}_{,j}\ .
\end{equation}
Lowering the contravariant index with $\eta_{ij}$ gives
\begin{equation}\label{50}
p_{i,j}=\frac{1}{r}\{\eta_{ij}+p_i\,p_j-(1-r\,h_0)^{-1}v_i\,v_j\}\ .
\end{equation}
Using the basis 1--forms listed in (\ref{45})--(\ref{48}) we derive the following formulae which will prove useful later:
\begin{eqnarray}
p_{i,j}\,\vartheta^i_{(1)}\,\vartheta^j_{(1)}&=&-\frac{1}{r}=p_{i,j}\,\vartheta^i_{(2)}\,\vartheta^j_{(2)}\ ,\label{51} \\
p_{i,j}\,\vartheta^i_{(1)}\,\vartheta^j_{(2)}&=&0=p_{i,j}\,\vartheta^i_{(3)}\,\vartheta^j_{(3)}\ ,\label{52}\\
p_{i,j}\,\vartheta^i_{(4)}\,\vartheta^j_{(4)}&=&-\frac{h_0}{1-r\,h_0}\ ,\label{53}\\
p_{i,j}\,\vartheta^i_{(1)}\,\vartheta^j_{(4)}&=&0=p_{i,j}\,\vartheta^i_{(2)}\,\vartheta^j_{(4)}\ ,\label{54}
\end{eqnarray}
from which, in particular, we find
\begin{equation}\label{55}
p^i{}_{,i}=\frac{2-3\,r\,h_0}{r\,(1-r\,h_0)}=\frac{2}{r}-h_0+O( r ) \ ,
\end{equation}
with the latter holding for small values of $r$.

\section{Frequency ratio in flat space--time}\label{sec_freq_ratio_flat}

With $X^i=(X, Y, Z, T)$ and $x^i=(x, y, r, u)$ the parametric equations of an arbitrary time--like world line, with arc length $s$ as parameter along it, are
\begin{equation}\label{56}
X^i=X^i(s)\ \ \ \Leftrightarrow\ \ \ x^i=x^i(s)\ ,
\end{equation}
with
\begin{equation}\label{57}
\eta_{ij}\,\frac{dX^i}{ds}\frac{dX^j}{ds}=+1\ .
\end{equation}
Using (\ref{22}) and writing
\begin{equation}\label{58}
\frac{dx^i}{ds}=\frac{dx^i}{du}\,\frac{du}{ds}\ ,
\end{equation}
in effect changing from the parameter $s$ to the parameter $u$ along the arbitrary world line, we can write
\begin{equation}\label{59}
\frac{dX^i}{ds}=\frac{du}{ds}\,\left\{[1+(\vec a\cdot\vec p)\,r]\,v^i+W^{(\alpha)}\,\lambda^i_{(\alpha)}\right\}\ ,
\end{equation}
with
\begin{equation}\label{60}
W^{(\alpha)}=p^{(\alpha)}\,\frac{dr}{du}+r\,\frac{\partial p^{(\alpha)}}{\partial x}\,\frac{dx}{du}+r\,\frac{\partial p^{(\alpha)}}{\partial y}\,\frac{dy}{du}+r\,\omega^{(\alpha)(\beta)}\,p_{(\beta)}\ .
\end{equation}
Substituting (\ref{59}) into (\ref{57}) we obtain
\begin{equation}\label{61}
\left (\frac{ds}{du}\right )^2=(1+(\vec a\cdot\vec p)\,r)^2-|\vec W|^2\ ,
\end{equation}
with
\begin{equation}\label{62}
\vec W=\vec u+r\,(\vec\omega\times\vec p)\ ,
\end{equation}
and
\begin{equation}\label{63}
\vec u=\frac{dr}{du}\,\vec p+r\,\frac{dx}{du}\,\frac{\partial\vec p}{\partial x}+r\,\frac{dy}{du}\,\frac{\partial\vec p}{\partial y}\ .
\end{equation}
We see that, since $u$ is proper--time along the world line $r=0$,  $\vec u$ is the 3--velocity of the observer with world line (\ref{56}) relative to the observer with world line $r=0$. Hence (\ref{61}) can finally be written
\begin{eqnarray}
\left(\frac{ds}{du}\right)^2&=&1-|\vec u|^2+2\,\{\vec a\cdot \vec p-\vec u\cdot(\vec\omega\times\vec p)\}\,r \nonumber \\
&&+\{(\vec a\cdot\vec p)^2-|\vec\omega\times\vec p|^2\}\,r^2\ .\label{66}
\end{eqnarray}
This formula is, of course, exact (in particular it does not have any restriction on $r$). The equation (\ref{61}), or equivalently the equation (\ref{66}), can be compared directly to the results in \cite{Hehl:Ni:1990} and with eq.\ (22) in \cite{Puetzfeld:Obukhov:2018:1}.

\section{Frequency ratio in curved space--time}\label{sec_freq_ratio_curved}

We now consider a general curved space--time. So that the line element of this space--time specializes easily to (\ref{25}) when the space--time is flat we need to emphasize some geometrical aspects which are also present in the the form (\ref{25}) of the Minkowskian line element. Guided by equations (\ref{36}) and (\ref{39}) we choose a family of time--like hypersurfaces $r(x^i)={\rm constant}$ in this space--time with unit space--like normal
\begin{equation}\label{67}
p^i=\frac{dx^i}{dr}=-g^{ij}\,r_{,j}\ \ \ {\rm and}\ \ \ g_{ij}\,p^i\,p^j=-1\ .
\end{equation}
Taking $x^3=r$ as a coordinate and labeling the remaining coordinates $x^i=(x, y, r, u)$ we have from (\ref{67})
\begin{equation}\label{68}
g_{ij}\,\frac{dx^j}{dr}=-r_{,i}\ \ \Leftrightarrow\ \ g_{i3}=-\delta^3_i\ .
\end{equation} 
Hence four components of the metric tensor of the space--time are fixed. Using straightforward algebra the remaining six components can be expressed in terms of six functions $P, \alpha, \beta, a, b, c$ of the four coordinates $x, y, r, u$ in such a way that the line element of the space--time is given by
\begin{equation}\label{69}
ds^2=-(\vartheta^{(1)})^2-(\vartheta^{(2)})^2-(\vartheta^{(3)})^2+(\vartheta^{(4)})^2\ ,
\end{equation}
with 
\begin{eqnarray}
\vartheta^{(1)}&=&-\vartheta_{(1)i}\,dx^i \nonumber \\
&=&r\,P^{-1}(e^{\alpha}\cosh\beta\,dx+e^{-\alpha}\sinh\beta\,dy+a\,du),\label{70}\\
\vartheta^{(2)}&=&-\vartheta_{(2)i}\,dx^i \nonumber \\
&=&r\,P^{-1}(e^{\alpha}\sinh\beta\,dx+e^{-\alpha}\cosh\beta\,dy+b\,du),\label{71}\\
\vartheta^{(3)}&=&-\vartheta_{(3)i}\,dx^i=dr,\label{72}\\
\vartheta^{(4)}&=&\vartheta_{(4)i}\,dx^i=c\,du.\label{73}
\end{eqnarray}
We take $r=0$ in this space--time to be a time--like world line with $u$ as proper time along it. Then in the neighborhood of $r=0$ (i.e. for small values of $r$) we expand the functions of $x, y, r, u$ in (\ref{70})--(\ref{73}) in positive powers of $r$, with coefficients functions of $x, y, u$, in such a way that the line element (\ref{69}) is a perturbation of the Minkowskian line element (\ref{25}). Clearly this involves taking
\begin{eqnarray}
P&=&P_0+O( r )\ ,\ \alpha=O( r )\ ,\ \beta=O( r )\ ,\ a=a_0+O( r )\ ,\nonumber\\
&&b=b_0+O( r )\ ,\ c=1-h_0\,r+O( r^2 )\ ,\label{74}
\end{eqnarray}
but we need to know the leading powers of $r$ in the $O( r )$--terms here. To find these we make use of (\ref{51})--(\ref{54}). With $p^i$ given by (\ref{67}), and using (\ref{70})--(\ref{73}), and denoting by a semicolon covariant differentiation with respect to the Riemannian connection associated with the metric tensor $g_{ij}$ given by the line element (\ref{69}) we start by recording that
\begin{eqnarray}
\vartheta^i_{(1)}&=&r^{-1}P\,(e^{-\alpha}\cosh\beta, -e^{\alpha}\sinh\beta, 0, 0)\ ,\label{75}\\
\vartheta^i_{(2)}&=&r^{-1}P\,(-e^{-\alpha}\sinh\beta, e^{\alpha}\cosh\beta, 0, 0)\ ,\label{76}\\
\vartheta^i_{(3)}&=&(0, 0, 1, 0)\ ,\label{77}\\
\vartheta^i_{(4)}&=&c^{-1}\Big(-e^{-\alpha}\{a\,\cosh\beta-b\,\sinh\beta\}, \nonumber \\
&&e^{\alpha}\{a\,\sinh\beta-b\,\cosh\beta\}, 0, 1\Big)\ ,\label{78}\end{eqnarray}
and thus
\begin{eqnarray}
p_{i;j}\,\vartheta^i_{(1)}\,\vartheta^j_{(1)}&-&p_{i;j}\,\vartheta^i_{(2)}\,\vartheta^j_{(2)}=-2\,\frac{\partial\alpha}{\partial r}\,\cosh 2\,\beta\ ,\label{79} \\
p_{i;j}\,\vartheta^i_{(1)}\,\vartheta^j_{(2)}&=&-\frac{\partial\beta}{\partial r}\  ,\label{80} \\
p_{i;j}\,\vartheta^i_{(1)}\,\vartheta^j_{(1)}&+&p_{i;j}\,\vartheta^i_{(2)}\,\vartheta^j_{(2)}=-2\frac{\partial}{\partial r}\log(r\,P^{-1})\ ,\label{81} \\
p_{i;j}\,\vartheta^i_{(4)}\,\vartheta^j_{(4)}&=&\frac{\partial}{\partial r}\log c\ ,\label{82} \\
p_{i;j}\,\vartheta^i_{(1)}\,\vartheta^j_{(4)}&=&\frac{1}{2}r\,(P\,c)^{-1}\left\{-\frac{\partial a}{\partial r}+(a\,\cosh 2\,\beta\right. \nonumber\\
&&\left.-b\,\sinh 2\,\beta)\frac{\partial\alpha}{\partial r}+b\,\frac{\partial\beta}{\partial r}\right\}\ ,\label{83}\\
p_{i;j}\,\vartheta^i_{(2)}\,\vartheta^j_{(4)}&=&\frac{1}{2}r\,(P\,c)^{-1}\left\{-\frac{\partial b}{\partial r}+(a\,\sinh 2\,\beta \right.\nonumber\\
&&\left.-b\,\cosh 2\,\beta)\frac{\partial\alpha}{\partial r}+a\,\frac{\partial\beta}{\partial r}\right\}\ .\label{84}
\end{eqnarray}

From (\ref{51}) perturbed for small values of $r$ we require the left hand sides of (\ref{79}) and (\ref{80}) to be small of order $r$ and this is achieved with
\begin{eqnarray}
\alpha&=&\alpha_2(x, y, u)\,r^2+O(r^3)\ ,\nonumber\\ 
\beta&=&\beta_2(x, y, u)\,r^2+O(r^3)\ .\label{85}
\end{eqnarray}
From (\ref{51}) we require the left hand side of (\ref{81}) to have the form $-2\,r^{-1}+O( r )$ and this is achieved with
\begin{equation}
P=P_0\{1+q_2(x, y, u)\,r^2+O(r^3)\}\ .\label{86}
\end{equation}
Next from (\ref{53}) the left hand side of (\ref{82}) should have the form $-h_0+O( r )$ and this occurs if 
\begin{equation}
c=1-h_0\,r+c_2(x, y, u)\,r^2+O(r^3)\ .\label{87}
\end{equation}
Finally from (\ref{54}) with $a=a_0+O( r )$ and $b=b_0+O( r )$ we now have from (\ref{83}) and (\ref{84}):
\begin{eqnarray}
p_{i;j}\,\vartheta^i_{(1)}\,\vartheta^j_{(4)}&=&-\frac{1}{2}r\,P_0^{-1}\frac{\partial a}{\partial r}+O(r^2)\ ,\label{88}\\
p_{i;j}\,\vartheta^i_{(2)}\,\vartheta^j_{(4)}&=&-\frac{1}{2}r\,P_0^{-1}\frac{\partial b}{\partial r}+O(r^2)\ ,\label{89}\end{eqnarray}
and to have the right hand sides of these $O( r )$ we take 
\begin{eqnarray}
a&=&a_0+a_1(x, y, u)\,r+O(r^2)\ ,\nonumber \\
b&=&b_0+b_1(x, y, u)\,r+O(r^2)\ .\label{90}
\end{eqnarray}
The components of the Riemann curvature tensor calculated on the world line $r=0$, and expressed on the orthonormal tetrad $\lambda^i_{(a)}$ with $a=1, 2, 3, 4$ defined by (\ref{7})--(\ref{11}), are denoted
\begin{equation}\label{91}
R_{(a)(b)(c)(d)}=R_{ijkl}\,\lambda^i_{(a)}\,\lambda^j_{(b)}\,\lambda^k_{(c)}\,\lambda^l_{(d)}.
\end{equation}
Calculating the Riemann tensor of the space--time evaluated on $r=0$ allows us to determine the functions $\alpha_2, \beta_2, q_2, c_2, a_1, b_1$ appearing in (\ref{85}), (\ref{86}), (\ref{87}) and (\ref{90}) in terms of the tetrad components (\ref{91}). We find the following expressions for these functions of $x, y, u$:
\begin{eqnarray}
\alpha_2&=&\frac{1}{6}P_0^2R_{(\alpha)(\beta)(\gamma)(\sigma)}\, \frac{\partial p^{(\alpha)}}{\partial x}\,p^{(\beta)}\,\frac{\partial p^{(\gamma)}}{\partial x}\,p^{(\sigma)} \nonumber\\
&&-\frac{1}{12}R_{(\alpha)(\beta)(\alpha)(\sigma)}\,p^{(\beta)}\,p^{(\sigma)}\ , \nonumber\\
&=&-\frac{1}{6}P_0^2R_{(\alpha)(\beta)(\gamma)(\sigma)}\,\frac{\partial p^{(\alpha)}}{\partial y}\,p^{(\beta)}\,\frac{\partial p^{(\gamma)}}{\partial y}\,p^{(\sigma)} \nonumber \\
&&+\frac{1}{12}R_{(\alpha)(\beta)(\alpha)(\sigma)}\,p^{(\beta)}\,p^{(\sigma)}\ ,\label{92}
\end{eqnarray}
with the second equality following from the use of (\ref{21}),

\begin{eqnarray}
\beta_2&=&\frac{1}{6}P_0^2R_{(\alpha)(\beta)(\gamma)(\sigma)}\,\frac{\partial p^{(\alpha)}}{\partial x}\,p^{(\beta)}\,\frac{\partial p^{(\gamma)}}{\partial y}\,p^{(\sigma)}\ ,\label{93} \\
q_2&=&-\frac{1}{12}\,R_{(\alpha)(\beta)(\alpha)(\sigma)}\,p^{(\beta)}\,p^{(\sigma)}\ ,\label{94}\\
c_2&=&-\frac{1}{2}\,R_{(\alpha)(4)(\beta)(4)}\,p^{(\alpha)}\,p^{(\beta)}\ ,\label{95}\\
a_1&=&-\frac{2}{3}\,P_0^2R_{(\alpha)(4)(\beta)(\gamma)}\,p^{(\alpha)}\,\frac{\partial p^{(\beta)}}{\partial x}\,p^{(\gamma)}\ ,\label{96}\\
b_1&=&-\frac{2}{3}\,P_0^2R_{(\alpha)(4)(\beta)(\gamma)}\,p^{(\alpha)}\,\frac{\partial p^{(\beta)}}{\partial y}\,p^{(\gamma)}\ .\label{97}
\end{eqnarray}
Fifteen equations which these functions satisfy are listed in the appendix \ref{sec_consistency_curvature} and can be verified directly using (\ref{21})--(\ref{21c}). When the functions are substituted into the 1-forms (\ref{70})--(\ref{73}) the line element (\ref{69}) is given, in the coordinates $x^i=(x, y, r, u)$ with $i=1, 2, 3, 4$ as $ds^2=g_{ij}\,dx^i\,dx^j$ with
\begin{eqnarray}
g_{11}&=&-r^2P_0^{-2}\{1+2\,(\alpha_2-q_2)\,r^2+O(r^3)\}\ ,\label{e1}\\
g_{22}&=&-r^2P_0^{-2}\{1-2\,(\alpha_2+q_2)\,r^2+O(r^3)\}\ ,\label{e2}\\
g_{12}&=&-r^2P_0^{-2}\{2\,\beta_2\,r^2+O(r^3)\}\ ,\label{e3}\\
g_{33}&=&-1\ \ (g_{31}=g_{32}=g_{34}=0)\ ,\label{e4}\\
g_{14}&=&-r^2\,P_0^{-2}\{a_0+a_1\,r+O(r^2)\}\ ,\label{e5}\\
g_{24}&=&-r^2\,P_0^{-2}\{b_0+b_1\,r+O(r^2)\}\ ,\label{e6}\\
g_{44}&=&\{1+(\vec a\cdot\vec p)\,r\}^2-|\vec\omega\times\vec p|^2r^2\nonumber \\
&&+2\,c_2\,r^2+O(r^3)\ ,\label{e7}\end{eqnarray}
with $P_0$ given by (\ref{17}).

If $x^i=x^i(s)$ with $x^i=(x, y, r, u)$ is an arbitrary time--like world line in the neighborhood of $r=0$ with $s$ proper time along it then, for small values of $r$ and using the 
line element (\ref{69}) the formula (\ref{66}) is modified to read
\begin{widetext}
\begin{eqnarray}
\left (\frac{ds}{du}\right )^2&=&1-|\vec u|^2+2\,\{\vec a\cdot \vec p-\vec u\cdot(\vec\omega\times\vec p)\}\,r+\{(\vec a\cdot\vec p)^2-|\vec\omega\times\vec p|^2\}\,r^2+2\,c_2\,r^2-2\,a_1\,r^3\,P_0^{-2}\,\frac{dx}{du}-2\,b_1\,r^3\,P_0^{-2}\,\frac{dy}{du}\nonumber\\
&&-4\,\beta_2\,r^4P_0^{-2}\frac{dx}{du}\,\frac{dy}{du}-2\,(\alpha_2-q_2)\,r^4P_0^{-2}\left (\frac{dx}{du}\right )^2+2\,(\alpha_2+q_2)r^4P_0^{-2}\left (\frac{dy}{du}\right )^2+O(r^3)\ .\nonumber\\\label{98}
\end{eqnarray}
Here $c_2$ is given by (\ref{95}). Using (\ref{96}) and (\ref{97}) we have
\begin{eqnarray}
a_1\,r^3P_0^{-2}\frac{dx}{du}+b_1\,r^3P_0^{-2}\frac{dy}{du}&=&-\frac{2}{3}r^2R_{(\alpha)(4)(\beta)(\gamma)}\,p^{(\alpha)}\,\Biggl (r\,\frac{dx}{du}\frac{\partial p^{(\beta)}}{\partial x}+
r\,\frac{dy}{du}\,\frac{\partial p^{(\beta)}}{\partial y}\Biggr )p^{(\gamma)}\ \nonumber\\
&=&-\frac{2}{3}\,r^2R_{(\alpha)(4)(\beta)(\gamma)}\,p^{(\alpha)}\,u^{(\beta)}\,p^{(\gamma)}\ ,\label{99}
\end{eqnarray}
using (\ref{63}). Next using (\ref{92})--(\ref{94}) and (\ref{63}) again we have
\begin{eqnarray}
&&-4\,\beta_2\,r^4P_0^{-2}\frac{dx}{du}\,\frac{dy}{du}-2\,(\alpha_2-q_2)\,r^4P_0^{-2}\left (\frac{dx}{du}\right )^2+2\,(\alpha_2+q_2)r^4P_0^{-2}\left (\frac{dy}{du}\right )^2\nonumber\\
&=&-\frac{1}{3}r^2R_{(\alpha)(\beta)(\gamma)(\sigma)}\,p^{(\beta)}\,p^{(\sigma)}\left (r\,\frac{dx}{du}\frac{\partial p^{(\alpha)}}{\partial x}+r\,\frac{dy}{du}\frac{\partial p^{(\alpha)}}{\partial y}\right )\left (r\,\frac{dx}{du}\frac{\partial p^{(\gamma)}}{\partial x}+r\,\frac{dy}{du}\frac{\partial p^{(\gamma)}}{\partial y}\right )\ \nonumber\\
&=& -\frac{1}{3}r^2R_{(\alpha)(\beta)(\gamma)(\sigma)}\,u^{(\alpha)}\,p^{(\beta)}\,u^{(\gamma)}\,p^{(\sigma)}\ .\label{101}
\end{eqnarray}
Substituting (\ref{99}) and (\ref{101}) into (\ref{98}) results in 
\begin{eqnarray}
\left (\frac{ds}{du}\right )^2&=&1-|\vec u|^2+2\,\{(\vec a\cdot \vec p)-\vec u\cdot(\vec\omega\times\vec p)\}\,r+\Biggl\{(\vec a\cdot\vec p)^2-|\vec\omega\times\vec p|^2-R_{(\alpha)(4)(\beta)(4)}\,p^{(\alpha)}\,p^{(\beta)}\nonumber\\
&&+\frac{4}{3}R_{(\alpha)(4)(\beta)(\gamma)}\,p^{(\alpha)}\,u^{(\beta)}\,p^{(\gamma)}-\frac{1}{3}R_{(\alpha)(\beta)(\gamma)(\sigma)}\,u^{(\alpha)}\,p^{(\beta)}\,u^{(\gamma)}\,p^{(\sigma)}\Biggr\}\,r^2+O(r^3)\ .\label{102}
\end{eqnarray}
\end{widetext}

\section{Plane Gravitational Waves I}\label{sec_plane_waves_1}

As a particularly simple illustration of the treatment of curvature above we consider the exact solution of Einstein's vacuum field equations which provides a space--time model of the gravitational field of plane gravitational waves. This well known solution is given by the line element
\begin{equation}\label{103}
ds^2=-dX^2-dY^2-dZ^2+dT^2+2\,H\,(dT-dZ)^2\ ,
\end{equation}
with
\begin{equation}\label{104}
H=a(T-Z)\,(X^2-Y^2)+2\,b(T-Z)\,X\,Y\ .
\end{equation}
A more general form for $H$, preserving the key properties for plane waves, namely, that $H$ is a harmonic function in $X, Y$ and the corresponding curvature tensor components are functions of $T-Z$ only, is required in section \ref{sec_radial_waves} below.

The histories of the plane wave fronts in the space--time with line element (\ref{103}) are the null hyperplanes
\begin{equation}\label{105}
T-Z={\rm constant}\ .
\end{equation}
The waves have two degrees of freedom of polarization reflected in the presence of the two arbitrary functions $a(T-Z)$ and $b(T-Z)$ and, in addition, their arbitrariness represents the freedom to choose the profile of the waves. In the coordinates $X^i=(X, Y, Z, T)$ the non--vanishing components of the Riemann curvature tensor are
\begin{eqnarray}
R_{1414}&=&-R_{2424}=R_{1313}=-R_{2323}=-R_{1413}\nonumber \\
&=&R_{2423}=-2\,a(T-Z)\ ,\label{106}
\end{eqnarray}
and
\begin{equation}\label{107}
R_{1424}=-R_{1423}=R_{1323}=-R_{1324}=-2\,b(T-Z)\ .
\end{equation}
From these it is clear that 
\begin{equation}\label{108}
R_{ijkm}\,k^m=0\ \ {\rm with}\ \ k^m=(0, 0, 1, 1)\ ,
\end{equation}
and so the curvature tensor is type N (purely radiative) in the Petrov classification with degenerate principal null direction $k^i$. The null vector field $k^i$ is covariantly constant and its expansion--free, twist--free and shear--free geodesic integral curves generate the null hyperplanes (\ref{105}). 

From (\ref{103}) and (\ref{104}) we see immediately that the coordinate $T$ is the arc length along the time--like world line $X=Y=Z=0$. The parametric equations of an arbitrary time--like world line in the space--time with line element (\ref{103}), with arc length $s$ along it, are $X^i=X^i(s)$ with
\begin{eqnarray}
&&-\left (\frac{dX}{ds}\right )^2-\left (\frac{dY}{ds}\right )^2-\left (\frac{dZ}{ds}\right )^2+\left (\frac{dT}{ds}\right )^2\nonumber\\
&&+2\,H\,\left (\frac{dT}{ds}-\frac{dZ}{ds}\right )^2=+1 \ .\label{109}
\end{eqnarray}
Using
\begin{equation}\label{110}
\vec u=(u^1,u^2,u^3)=\left (\frac{dX}{dT}, \frac{dY}{dT}, \frac{dZ}{dT}\right) ,
\end{equation}
which is the 3--velocity of the observer with world line $X^i=X^i(s)$ measured by the observer with world line $X=Y=Z=0$, we can rewrite (\ref{109}) in the form
\begin{equation}\label{111}
\left (\frac{ds}{dT}\right )^2=1-|\vec u|^2+2\,H\,\left (1-u^3\right )^2\ .
\end{equation}
Using (\ref{104}), (\ref{106}) and (\ref{107}) we see that 
\begin{equation}\label{112}
R_{A4B4}\,X^A\,X^B=-2\,H\quad \quad(A,B=1,2),
\end{equation}
with $X^A=(X, Y)$. On account of the simplicity of the Riemann tensor (in particular that it has only two independent components) all of the information contained in it can be extracted using the observer with world line $X=Y=Z=0$ and observers with world lines $Z={\rm constant}$. The ratio of arc lengths or proper--times along such world lines is, by (\ref{111}) and (\ref{112}), 
\begin{equation}\label{113}
\left (\frac{ds}{dT}\right )^2=1-u^A\,u^A-R_{A4B4}\,X^A\,X^B\ .
\end{equation}
We note that in general the final term in (\ref{111}) can be written 
\begin{eqnarray}\label{114}
2\,H\,\left (1-u^3\right )^2&=&-R_{A4B4}\,X^A\,X^B-2\,R_{A4B3}\,X^A\,X^B\,u^3\nonumber\\
&&-R_{A3B3}\,X^A\,X^B\,(u^3)^2\ .
\end{eqnarray}
However this contains no more information on the Riemann tensor than the final term in (\ref{113}) since
\begin{eqnarray}
R_{A4B3}\,X^A\,X^B&=&-R_{A3B3}\,X^A\,X^B=2\,H\nonumber\\
&=&-R_{A4B4}\,X^A\,X^B\ .\label{115}
\end{eqnarray} 
 
\section{Plane Gravitational Waves II}\label{sec_plane_waves_2}

The function $H$ in (\ref{103}) and (\ref{104}) has the property that it vanishes on the world line $X=Y=Z=0$. Its essential analytical properties are that the vacuum field equations require it to be a harmonic function,
\begin{equation}\label{116}
H_{XX}+H_{YY}=0\ ,
\end{equation}
with the subscripts denoting partial derivatives, and the curvature tensor components must be functions of $T-Z$ so that 
\begin{equation}\label{117}
H_{XX}-H_{YY}=4\,a(T-Z)\ \ \ {\rm and}\ \ \ H_{XY}=2\,b(T-Z)\ .
\end{equation}
Hence we can have it vanish on the arbitrary time--like world line $X^i=w^i(u)$ by taking it to be
\begin{multline}
H=a(T-Z)\ \{(X-w^1(T-Z))^2-(Y-w^2(T-Z))^2\}\nonumber\\
+2\,b(T-Z)\,(X-w^1(T-Z))(Y-w^2(T-Z))\ .\label{118}
\end{multline}
With $R_{ijkl}$ given by (\ref{106}) and (\ref{107}) we can write this as (again with capital indices taking values 1, 2)
\begin{equation}\label{119}
2H=-R_{A4B4}(T-Z)[X^A-w^A(T-Z)][X^B-w^B(T-Z)].
\end{equation} 
We now make the coordinate transformation
\begin{eqnarray}
X^i&=&w^i+r\,p^i+\frac{1}{3}r^3(p^4-p^3)(v^4-v^3)\nonumber\\
&&\times R_{A4B4}\,p^A\,p^B\,v^i+O(r^4)\ ,\label{120}
\end{eqnarray}
which generalises (\ref{33}) for small values of $r$ and therefore applies in the neighborhood of the time--like world line $r=0$. The effect of this on the line element (\ref{103}) with $H$ given by (\ref{119}) is to transform it into
\begin{eqnarray}
ds^2&=&-r^2P_0^{-2}\{(dx+a_0\,du)^2+(dy+b_0\,du)^2\}-dr^2\nonumber\\
&&+\Big\{1-2\,h_0r+h_0^2r^2-r^2(v^4-v^3)^2\nonumber\\ 
&&\times R_{A4B4}\,p^A\,p^B\Big\}\,du^2\ ,\label{121}
\end{eqnarray}
neglecting $O(r^3)$--terms. Here $P_0, a_0, b_0, h_0$ are given by (\ref{17}) and (\ref{32_1}). This form of the line element of the space--time model of the gravitational field of plane gravitational waves is in the form of line element discussed in section 4. To effect a closer comparison we note that
\begin{eqnarray}
R_{(\alpha)(4)(\beta)(4)}\,p^{(\alpha)}\,p^{(\beta)}&=&R_{ijkl}\,p^i\,v^j\,p^k\,v^l\nonumber\\ 
&=&R_{A4B4}\,U^A\,U^B\ ,\label{122}
\end{eqnarray}
with
\begin{equation}\label{123}
U^A=(v^4-v^3)\,p^A-(p^4-p^3)\,v^A\ .
\end{equation}
Hence if $v^A=0$, so that the time--like world line $r=0$ is the history of an observer accelerating in the direction of propagation of the gravitational waves (the $Z$--direction), then in this case (\ref{102}) simplifies to 
\begin{eqnarray}
&&\left (\frac{ds}{du}\right )^2=1-|\vec u|^2+2\,\{(\vec a\cdot \vec p)-\vec u\cdot(\vec\omega\times\vec p)\}\,r\nonumber\\ 
&&+\Biggl\{(\vec a\cdot\vec p)^2-|\vec\omega\times\vec p|^2-R_{(\alpha)(4)(\beta)(4)}\,p^{(\alpha)}\,p^{(\beta)}\Biggr\}\,r^2+O(r^3)\ . \nonumber\\ 
&&\label{124}
\end{eqnarray}
The origin of the coordinate transformation (\ref{120}) is to start with the line element 
\begin{eqnarray}
ds^2&=&\eta_{ij}\,dX^i\,dX^j+2\,H\,(dT-dZ)^2\ \nonumber\\
&=&\eta_{ij}\,dX^i\,dX^j-R_{A4B4}(T-Z)\,\Big\{X^A-w^A(T-Z)\Big\}\nonumber\\
&&\times \Big\{X^B-w^B(T-Z)\Big\}(dT-dZ)^2\ .\label{125}
\end{eqnarray} 
Now in the final term here make the transformation (\ref{33}). This involves
\begin{eqnarray}
&&T-Z=w^4-w^3+r\,(p^4-p^3)\Rightarrow\nonumber\\ 
&& dT-dZ=(v^4-v^3)du+(p^4-p^3)\,dr+O( r ),\label{126}
\end{eqnarray}
and
\begin{eqnarray}
&&R_{A4B4}(T-Z)\,(X^A-w^A(T-Z))\,(X^B-w^B(T-Z))\nonumber\\
&&=r^2R_{A4B4}(u)\,p^A\,p^B+O(r^3)\ .\label{127}
\end{eqnarray}
Hence the final term in the line element (\ref{125}) reads
\begin{eqnarray}
&&-r^2R_{A4B4}\,p^A\,p^B\Big\{(v^4-v^3)^2\,du^2\nonumber\\
&&+2\,(v^4-v^3)(p^4-p^3)\,du\,dr\Big\}+O(r^3)\ .\label{128}
\end{eqnarray}
Now to calculate $\eta_{ij}\,dX^i\,dX^j$ we modify the transformation (\ref{33}) to (\ref{120}) in order to cancel the $du\,dr$--term in (\ref{128}) when everything is substituted into the line element (\ref{125}). From (\ref{120}) it follows that
\begin{eqnarray}
dX^i&=&(v^i+r\,\frac{\partial p^i}{\partial u})du+r\,\frac{\partial p^i}{\partial x}\,dx+r\,\frac{\partial p^i}{\partial y}\,dy\nonumber\\
&&+\left\{p^i+r^2v^i\,(v^4-v^3)(p^4-p^3)R_{A4B4}\,p^A\,p^B\right\}dr\nonumber\\
&&+O(r^3)\ .\label{19_2}
\end{eqnarray}
Since $v^i$ and $p^i$ are orthogonal the only surviving Riemann tensor term in $\eta_{ij}\,dX^i\,dX^j$ is $2\,r^2(v^4-v^3)(p^4-p^3)R_{A4B4}\,p^A\,p^B\,du\,dr$ (neglecting $O(r^3)$--terms) and so when $\eta_{ij}\,dX^i\,dX^j$ is added to (\ref{128}) now the result is the line element (\ref{121}).

\section{Waves Moving Radially Relative to $r=0$}\label{sec_radial_waves}

The plane gravitational waves have the property that their propagation direction in space--time is covariantly constant. Hence their propagation direction in space--time is, in particular, \emph{non--expanding}. Arguably the simplest example of gravitational waves for which the propagation direction in space--time is not covariantly constant \emph{and is expanding} are waves moving radially with respect to the observer with world line $r=0$ in the present context. Such waves may, for example, be spherical fronted but the wave fronts cannot be centered on the observer with world line $r=0$ since that would result in the Riemann curvature tensor being singular on $r=0$ which emphatically is \emph{not} the case here. It follows from (\ref{47}) and (\ref{72}) that the 3--direction is the radial direction relative to the world line $r=0$. We thus consider gravitational waves whose propagation direction calculated on $r=0$ is given by the 1--form
\begin{equation}\label{r1}
k_{(a)}\,\vartheta^{(a)}=-\vartheta^{(3)}+\vartheta^{(4)}\ \ \Leftrightarrow\ \ k^{(a)}=(0, 0, 1, 1)\ .
\end{equation}
Thus for small values of $r$,
\begin{equation}\label{r2}
k_{(a)}\,\vartheta^{(a)}=\{-r_{,i}+(1-r\,h_0)\,u_{,i}+O(r^2)\}\,dX^i=k_i\,dX^i\ ,
\end{equation}
and, using (\ref{35}) and (\ref{36}), we can write
\begin{eqnarray}
k_i&=&-r_{,i}+(1-r\,h_0)\,u_{,i}+O(r^2) \nonumber \\
&=&p_i+v_i+O(r^2)\ \ (\Rightarrow\ k^i\,k_i=O(r^2))\ ,\label{r4}
\end{eqnarray}
and so the light--like propagation direction calculated \emph{on} $r=0$ is $k^i=p^i+v^i$. The vacuum field equations
\begin{equation}\label{r5}
R_{(a)(b)}=-R_{(\alpha)(a)(b)(\alpha)}+R_{(4)(a)(b)(4)}=0\ ,
\end{equation}
and the radiative conditions on the Riemann tensor (that the Riemann tensor be type N in the Petrov classification with $k^{(a)}$ as degenerate principal null direction) 
\begin{equation}\label{r6}
R_{(a)(b)(c)(d)}\,k^{(d)}=R_{(a)(b)(c)(3)}+R_{(a)(b)(c)(4)}=0\ ,
\end{equation}
must be satisfied on $r=0$ for substitution into (\ref{102}). As a consequence of (\ref{r5}) and (\ref{r6}) there are only two independent non--vanishing components of the vacuum Riemann tensor calculated on $r=0$, namely, $R_{(1)(4)(1)(4)}=-R_{(2)(4)(2)(4)}$ and $R_{(1)(4)(2)(4)}$. All remaining non--vanishing curvature components are given in terms of these by
\begin{eqnarray}
R_{(1)(3)(1)(3)}&=&-R_{(2)(3)(2)(3)}=-R_{(1)(3)(1)(4)}\nonumber \\
&=&R_{(2)(3)(2)(4)}=R_{(1)(4)(1)(4)}\ ,\label{r7}
\end{eqnarray}
and
\begin{eqnarray}
R_{(1)(3)(2)(3)}&=&-R_{(1)(4)(2)(3)}=-R_{(2)(4)(1)(3)}\nonumber \\
&=&R_{(1)(4)(2)(4)}\ .\label{r8}
\end{eqnarray}
When these are substituted into the Riemann tensor terms in (\ref{102}) we find that 
\begin{eqnarray}
R_{(\alpha)(4)(\beta)(4)}\,p^{(\alpha)}\,&p^{(\beta)}&=R_{(A)(4)(B)(4)}\,p^{(A)}\,p^{(B)}\ , \nonumber \\ \label{r9}\\
R_{(\alpha)(4)(\beta)(\gamma)}\,p^{(\alpha)}\,u^{(\beta)}\,&p^{(\gamma)}&\nonumber \\ 
=R_{(A)(4)(B)(4)}\,\{&u^{(3)}&\,p^{(A)}-u^{(A)}\,p^{(3)}\}\,p^{(B)}\ , \label{r10}
\end{eqnarray}
and
\begin{eqnarray}
&&R_{(\alpha)(\beta)(\gamma)(\sigma)}\,u^{(\alpha)}\,p^{(\beta)}\,u^{(\gamma)}\,p^{(\sigma)}=R_{(A)(4)(B)(4)} \nonumber \\
 && \times \{u^{(A)}\,p^{(3)}-u^{(3)}\,p^{(A)}\} \{u^{(B)}\,p^{(3)}-u^{(3)}\,p^{(B)}\},\label{r11}
\end{eqnarray}
where capital letters take values 1, 2.

Substituting (\ref{r9})-(\ref{r11}) into (\ref{102}) we find 
\begin{eqnarray}
\left (\frac{ds}{du}\right )^2&&=1-|\vec u|^2+2\,\{(\vec a\cdot \vec p)-\vec u\cdot(\vec\omega\times\vec p)\}\,r \nonumber \\
&& +\Biggl\{(\vec a\cdot\vec p)^2-|\vec\omega\times\vec p|^2-R_{(A)(4)(B)(4)}\,p^{(A)}\,p^{(B)}\nonumber \\
&&+\frac{4}{3}R_{(A)(4)(B)(4)}\,\{u^{(3)}\,p^{(A)}-u^{(A)}\,p^{(3)}\}\,p^{(B)}\nonumber\\
&&-\frac{1}{3}R_{(A)(4)(B)(4)}\{u^{(A)}\,p^{(3)}-u^{(3)}\,p^{(A)}\} \nonumber \\
&& \times \{u^{(B)}\,p^{(3)}-u^{(3)}\,p^{(B)}\}\Biggr\}\,r^2+O(r^3)\ .\label{102_rewritten}
\end{eqnarray}
It is interesting to note that while $k^i$ given by (\ref{r4}) when $r=0$ is the propagation direction of the radial gravitational waves relative to the observer with world line $r=0$ it cannot be the propagation direction of gravitational waves in the neighborhood of $r=0$ (i.e.\ for small, non--zero, values of $r$). The reason for this is because the Goldberg--Sachs \cite{Goldberg:Sachs:1962} theorem requires the propagation direction in space--time of gravitational waves propagating in a vacuum to be geodesic and shear--free. Using (\ref{35}) and (\ref{50}) we have
\begin{equation}\label{r12}
k_{i,j}=\frac{1}{r}(\eta_{ij}+p_i\,p_j-v_i\,v_j)-h_0\,v_i\,v_j+a_i\,v_j+O( r )\ ,
\end{equation}
from which we conclude that
\begin{equation}\label{r13}
k_{i,j}\,k^j=-h_0\,v_i+O( r )\ ,
\end{equation}
and so $k^i$ is not even approximately geodesic for small $r$ if $a^i\neq 0$ (i.e. if $r=0$ is not a time--like geodesic).
However we can construct an approximately null vector field $K^i$ in the neighborhood of $r=0$, which coincides with $k^i$ on $r=0$, and which is approximately geodesic and shear--free. Such a vector field is given by
\begin{equation}\label{r14}
K^i=p^i+v^i-\frac{1}{2}\{a^i+h_0\,p^i\}\,r+O(r^2)\ \ \Rightarrow\ \ K^i\,K_i=O(r^2)\ .
\end{equation}
When differentiating this with respect to $X^i$ using (\ref{35}), (\ref{36}) and (\ref{50}) it is useful to note that 
the partial derivative of $h_0=a_i\,p^i$ reads
\begin{equation}\label{r15}
\frac{\partial h_0}{\partial X^i}=\frac{1}{r}(a_i+h_0\,p_i)+O(r^0)\ .
\end{equation}
In particular we calculate that
\begin{equation}\label{r16}
K_{i,j}+K_{j,i}=\lambda\,\eta_{ij}+\xi_i\,K_j+\xi_j\,K_i+O( r )\ ,
\end{equation}
with
\begin{equation}\label{r17}
\lambda=\frac{2}{r}-h_0+O( r )\ ,
\end{equation}
and 
\begin{equation}\label{r18}
\xi_i=\frac{1}{r}(p_i-v_i)+\frac{1}{2}(a_i-h_0\,v_i)+O( r )\ .
\end{equation}
The appearance of the algebraic form of the right hand side of (\ref{r16}) ensures that $K^i$ is geodesic and shear--free in the neighborhood of $r=0$ (i.e.\ $K^i$ is geodesic and shear--free if $O( r )$--terms are neglected). This characterization of ``geodesic and shear--free" is due to Robinson and Trautman \cite{Robinson:Trautman:1983}. It is useful for discussing these geometrical properties when, (a) not using a null tetrad and (b) not assuming an affine parameter along the integral curves of the null vector field. We note 
in particular that it follows from (\ref{r16}) that 
\begin{equation}\label{r19}
K_{i,j}\,K^j=-h_0\,K_i+O( r )\ ,
\end{equation}
demonstrating that $K^i$ is approximately geodesic (without an affine parameter if $h_0\neq 0$).

\section{Clock compass}\label{sec_clock_compass}

In the following, the general idea is to use suitably prepared set of clocks to determine all components of the gravitational field. The goal is to express all parameters of the space--times under consideration by means of the measured frequency ratios between the clocks in a configuration. In analogy to the gravitational compass \cite{Szekeres:1965,Puetzfeld:Obukhov:2016:1}, we call such a clock configuration a ``gravitational clock compass'' \cite{Puetzfeld:Obukhov:2018:1}.

In contrast to the general procedure outlined in \cite{Puetzfeld:Obukhov:2018:1}, in which we worked out the minimal number clocks necessary for a measurement of all the components of the gravitational field, we now consider setups of clocks which allow for a determination of the properties of the special space--times introduced in the previous sections. 

In the following we are going to search for arrangements of $n$ clocks, at positions ${}^{(n)}\!p^{{\alpha}}$ w.r.t.\ the reference world line of the observer. In addition to the positions of the compass constituents, we may also make a choice for the velocity of the clocks w.r.t.\ the observer, denoted by ${}^{(m)}\!u^{{\alpha}}$ in the following. While possible in principle, and in particular covered by our general formalism, we are not going to allow for situations with additional accelerations or rotations.
 
\subsection{Plane gravitational waves}

The starting point is (\ref{124}), which is the measureable frequency ratio as a function $C=C(r, p^\alpha, u^\alpha, a^\alpha, \omega^{\alpha \beta}, R_{\alpha \beta \gamma \delta})$ of the quantities characterizing the state of motion as well as the space--time.

Assuming that all quantities but the gravitational field can be prescribed by the experimentalist, we can rearrange (\ref{124}) as follows:
\begin{eqnarray}
&&B(r, p^\alpha, u^\alpha, a^\alpha, \omega^{\alpha \beta}) =R_{(\alpha)(4)(\beta)(4)}\,p^{(\alpha)}\,p^{(\beta)}\ , \nonumber\\ 
&&\label{124_1}
\end{eqnarray}
where
\begin{eqnarray}
&&B(r, p^\alpha, u^\alpha, a^\alpha, \omega^{\alpha \beta}) :=  (\vec a\cdot\vec p)^2-|\vec\omega\times\vec p|^2   \nonumber\\ 
&&+\frac{2}{r}\Bigl\{(\vec a\cdot \vec p)-\vec u\cdot(\vec\omega\times\vec p)\Bigr\}+ \frac{1}{r^2} \Bigl( 1- C -|\vec u|^2\Bigr).\label{124_2}
\end{eqnarray}  

\begin{figure}
\begin{center}
\includegraphics[width=2cm,angle=-90]{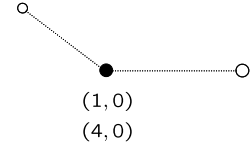}
\end{center}
\caption{\label{fig_1} Symbolical sketch of the explicit clock configuration which allows for a complete determination of the gravitational field (\ref{plane_sol_a_b}). In total 2 suitably prepared clocks (hollow circles) are needed to determine all curvature components. The observer is denoted by the black circle. We make use of the notation analogous to the one in \cite{Puetzfeld:Obukhov:2018:1}.}
\end{figure}

Employing the strategy from \cite{Puetzfeld:Obukhov:2016:1,Puetzfeld:Obukhov:2018:1}, we are now looking for a configuration of clocks, which allows for a determination of all components of the gravitational field in terms of the measured quantities $B$. By labeling different positions of the clocks by an additional index $(n)$ equation (\ref{124_1}) turns into the system
\begin{eqnarray}
&&{}^{(n)}\!B =R_{(\alpha)(4)(\beta)(4)}\,{}^{(n)}\!p^{(\alpha)}\,{}^{(n)}\!p^{(\beta)}\ , \label{124_3}
\end{eqnarray}
in which we suppressed all indices of quantities entering ${}^{(n)}\!B$ which are directly controlled by the experimentalist. Considering different choices for the positions ${}^{(n)}\!p^{\alpha}$, we notice that we end up with the constrained vacuum clock compass solution given in \cite[(114)--(119)]{Puetzfeld:Obukhov:2018:1}:
\begin{eqnarray}
01 : R_{(1)(4)(1)(4)}&=& {}^{(1)}B, \label{constrained_curvature_1} \\
02 : R_{(2)(4)(2)(4)}&=& {}^{(2)}B, \label{constrained_curvature_2} \\
03 : R_{(3)(4)(3)(4)}&=& {}^{(3)}B, \label{constrained_curvature_3} \\
04 : R_{(2)(4)(1)(4)}&=& \frac{1}{2} \left({}^{(4)}B - {}^{(1)}B - {}^{(2)}B\right), \label{constrained_curvature_4} \\
05 : R_{(3)(4)(2)(4)}&=& \frac{1}{2} \left({}^{(5)}B - {}^{(2)}B - {}^{(3)}B\right), \label{constrained_curvature_5} \\
06 : R_{(3)(4)(1)(4)}&=& \frac{1}{2} \left({}^{(6)}B - {}^{(1)}B - {}^{(3)}B\right). \label{constrained_curvature_6} 
\end{eqnarray} 
Of course in our case the situation is simplified even further due to (\ref{106}) and (\ref{107}). From the constrained system we can infer -- using the notation from  \cite{Puetzfeld:Obukhov:2018:1} -- that two clocks at positions 
\begin{eqnarray}
{}^{(1)}\!p^{{\alpha}}=\left(\begin{array}{c} 1 \\ 0\\ 0\\ \end{array} \right), \quad \quad
{}^{(4)}\!p^{{\alpha}}=\left(\begin{array}{c} 1\\ 1\\ 0\\ \end{array} \right), \label{positions_setup}
\end{eqnarray}
allow for a complete determination of the gravitational field, i.e.\ the functions $a$ and $b$ are given by
\begin{eqnarray}
a = -\frac{1}{2}\, {}^{(1)}\!B, \quad \quad b = - \frac{1}{4}\,{}^{(4)}\!B. \label{plane_sol_a_b}  
\end{eqnarray}
See figure \ref{fig_1} for a symbolical sketch of the solution. Note that the sketches of the clock configurations make use of a notation analogous to the one in \cite{Puetzfeld:Obukhov:2018:1}. The observer is indicated by a black circle, the prepared clocks are indicated by hollow circles. In contrast to the notation in (\ref{plane_sol_a_b}) -- in which all indices but the relevant position index $(n)$ are suppressed -- the second (velocity) index $(m)$ is explicitly given in figure \ref{fig_1} and set to $m=0$, indicating that the clocks in this configuration do not move w.r.t.\ to the observer. Furthermore, we note that the sketches were introduced in \cite{Puetzfeld:Obukhov:2018:1} to give a 2 dimensional visual representation of the solution. In particular they are designed for counting the number of clocks/measurements at a glance, they do not directly represent the 3 dimensional geometry of the measurement (we order hollow circles, corresponding to different positions $(n)$, starting at the three o'clock position, advancing counter clockwise in 45 degree angles depending on the position index $n$).

\subsection{Waves radial relative to $r=0$}

Following the same line of reasoning as in the case of plane gravitational waves, we use the definition for $B$ as given in (\ref{124_2}), however now we have a system of clocks at positions ${}^{(n)}p^{(\alpha)}$ moving with velocities ${}^{(m)}u^{(\alpha)}$, and we are left with the system
\begin{eqnarray}
  {}^{(n,m)}B&& = {}^{(n)}p^{(\alpha)}  {}^{(n)}p^{(\beta)} \Bigg(R_{(\alpha) (4) (\beta) (4)} - \frac{4}{3} R_{(\alpha) (4) (\gamma) (\beta)} \nonumber \\
	&& \times \, {}^{(m)}u^{(\gamma)} + \frac{1}{3} R_{ (\gamma) (\alpha) (\delta) (\beta) } \, {}^{(m)}u^{(\gamma)} {}^{(m)}u^{(\delta)} \Bigg). \label{radial_compass}
\end{eqnarray}

In vacuum, the general clock compass solution on the basis of (\ref{radial_compass}) was given in \cite{Puetzfeld:Obukhov:2018:1}. Taking into account the non-vanishing curvature components in the radial case as indicated in (\ref{r7}) and (\ref{r8}), one may infer several clock configurations which allow for a determination of the curvature components. 

One configuration coincides with the one already given in plane gravitational wave case, c.f.\ eq.\ (\ref{plane_sol_a_b}) and figure \ref{fig_1}. However, due to the more general nature of the compass equation (\ref{radial_compass}) one may now also construct configurations in which the clocks are in motion. We briefly mention here two possible configurations, i.e.\
\begin{eqnarray}
R_{(1)(4)(1)(4)} &=& 3 c_{11}^{-2} \,\, {}^{(3,1)}B, \label{radial_R1414_sol} \\
R_{(1)(4)(2)(4)} &=& \left(2 - \frac{8}{3} c_{33} + \frac{2}{3}c_{33}^2 \right)^{-1} \, {}^{(4,3)}B  \label{radial_R1424_sol}.
\end{eqnarray}
An alternative solution for the second curvature component is given by
\begin{eqnarray}
R_{(1)(4)(2)(4)} &=& \frac{3}{2 c_{41} c_{42}} \left\{{}^{(3,4)}B \right. \nonumber \\ 
&&\left. -\left[\left(\frac{c_{41}}{c_{11}}\right)^2 -\left(\frac{c_{42}}{c_{11}}\right)^2 \right] {}^{(3,1)}B   \right\}. \nonumber \\ \label{radial_R1424_alt_sol}
\end{eqnarray}
Here we used the same nomenclature for the positions and velocities as in \cite{Puetzfeld:Obukhov:2018:1}, i.e.
\begin{eqnarray}
{}^{(3)}\!p^{{\alpha}}=\left(\begin{array}{c} 0 \\ 0\\ 1\\ \end{array} \right), \quad {}^{(1)}\!u^{{\alpha}}=\left(\begin{array}{c} c_{11}\\ 0\\ 0\\ \end{array} \right),\nonumber \\
{}^{(3)}\!u^{{\alpha}}=\left(\begin{array}{c} 0\\ 0\\ c_{33}\\ \end{array} \right), \quad {}^{(4)}\!u^{{\alpha}}=\left(\begin{array}{c} c_{41}\\ c_{42}\\ 0\\ \end{array} \right). \label{positions_velocities_setup}
\end{eqnarray}

\begin{figure}[h]
\begin{center}
\includegraphics[width=3.1cm,angle=-90]{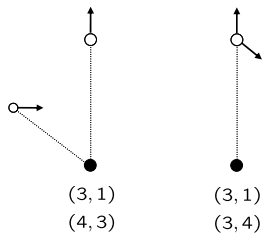}
\end{center}
\caption{\label{fig_2} Symbolical sketch of the two explicit clock configurations which allow for a complete determination of the gravitational field (\ref{radial_R1414_sol})-(\ref{radial_R1424_alt_sol}). In both cases two suitably prepared clocks (hollow circles) are needed to determine all curvature components. Again the observer is denoted by the black circle. We make use of the notation analogous to the one in \cite{Puetzfeld:Obukhov:2018:1}.}
\end{figure}

Symbolical sketches of the solutions (\ref{radial_R1414_sol})-(\ref{radial_R1424_alt_sol}) are given in figure \ref{fig_2}. Note that we order arrows, corresponding to different velocities $(m)$, starting at the twelve o'clock position, advancing clockwise in 45 degree angles depending on the velocity index $m$.

\section{Conclusions}\label{sec_conclusions}

In this work we presented an alternative derivation of the gravitational clock compass, previously proposed in \cite{Puetzfeld:Obukhov:2018:1,Puetzfeld:Obukhov:2018:2}, by means of the approximation technique developed in \cite{Hogan:Trautman:1987,Hogan:Futamase:Itoh:2008,Hogan:Asada:Futamase:2010}.

It should be emphasized that the derivation presented here starts from scratch, i.e.\ from first principles in flat space--time. It is reassuring to observe that the result regarding the general frequency ratio from \cite{Puetzfeld:Obukhov:2018:1} can, within the conventions used in the present work, be confirmed by the use of an independent approximation technique.

Building upon this result, we were able to specialize the general compass setup to two special types of space--times, describing plane gravitational waves and waves moving radially to an observer. 

It should be stressed that the focus of the present work differs somewhat from other works in the gravitational wave context, for here the main focus is on the general geometry of the clock configuration required for a complete field determination, and not the possible measurement of the wave character (profile). In contrast to classical works on (indirect) timing experiments like \cite{Detweiler:1979,Estabrook:Wahlquist:1975}, a clock compass relies on the direct frequency comparison of a suitably prepared set of local clocks. 

It is clear that the highly idealized situations of plane and spherically gravitational waves should be generalized in future works. Still they serve as a testbed and demonstrate the direct operational relevance of a clock compass. We hope to be able to extend them in future works to an approximate description of more general radiative space--times. A future goal would be the realization of an omnidirectional (tensorial) \cite{Forward:1971,Wagoner:Paik:1976,Paik:etal:2016} gravitational wave detector based on clocks.

\begin{acknowledgments}
This work was funded by the Deutsche Forschungsgemeinschaft (DFG, German Research Foundation) through the grant PU 461/1-2 -- project number 369402949 (D.P.). 
\end{acknowledgments}

\appendix

\section{Notations and conventions}\label{sec_notation}

Note that our conventions for labeling the space--time metric differs from the one in \cite{Puetzfeld:Obukhov:2018:1}. The signature is assumed to be $(-1,-1,-1,+1)$. Latin indices run from $1, \dots, 4$, and Greek indices from $1, \dots, 3$.

\begin{table}
\caption{\label{tab_symbols}Directory of symbols.}
\begin{ruledtabular}
\begin{tabular}{ll}
Symbol & Explanation\\
\hline
$ds$ & Line element \\
$g_{a b}$, $\eta_{a b}$ & Metric, flat metric\\
$\vartheta^a$ & Coframe\\
$\delta^a_b$ & Kronecker symbol \\
$x^{a}$, $X^{a}$ & Coordinates \\
$x,y$ & Stereographic coordinates\\
$u$, $s$ & Proper time \\
$P^a$& Space-like vectorfield\\
$R_{a b c}{}^d$ & Riemann curvature\\
$\lambda^a{}_{(b)}$ & Orthonormal tetrad \\
$x^i(s)$ & (Reference) world line\\
$v^a$& Velocity\\
$\omega^{ab}$ & Rotation\\
$a^\alpha$ & Acceleration\\
$C$ & Frequency ratio \\
$w^i$, $P$, $P_0$, $r$, $h_0$, $a$, $a_{0,1,2}$, & Auxiliary quantities\\
$b$, $b_{0,1,2}$, $c$, $c_0$, $q$, $q_2$, $\alpha$, $\alpha_2$, & \\
$\beta$, $\beta_2$, $H$, $U^A$, $B$ & \\
&\\
\hline
\multicolumn{2}{l}{{Operators}}\\
\hline
($\partial_i$,``\,$,$\,'') , ($\nabla_i$,``\,$;$\,'') & (Partial, covariant) derivative \\ 
``\,$\vec{\phantom{a}}$\,'' & 3d vector \\
``\,$\cdot$\,'' & 3d scalar product \\
``\,$\times$\,'' & 3d vector product \\
&\\
\end{tabular}
\end{ruledtabular}
\end{table}

\section{Consistency of Curvature Tensor Calculation}\label{sec_consistency_curvature}

The following equations are a spin--off from the calculations of the Riemann tensor and can be verified to be satisfied by $\alpha_2, \beta_2, q_2, c_2, a_1, b_1$ given 
in (\ref{92})--(\ref{97}) using (\ref{21})--(\ref{21c}):
\begin{widetext}
\begin{eqnarray}
\frac{1}{2}\left (\frac{\partial}{\partial x}(P_0^{-2}b_1)-\frac{\partial}{\partial y}(P_0^{-2}a_1)\right )&=&R_{(\alpha)(4)(\beta)(\gamma)}\,p^{(\alpha)}\,\frac{\partial p^{(\beta)}}{\partial x}\frac{\partial p^{(\gamma)}}{\partial y}\ ,\label{A1} \\
\frac{\partial c_2}{\partial x}&=&-R_{(\alpha)(4)(\beta)(4)}\,p^{(\alpha)}\,\frac{\partial p^{(\beta)}}{\partial x}\ ,\label{A2}\\
\frac{\partial c_2}{\partial y}&=&-R_{(\alpha)(4)(\beta)(4)}\,p^{(\alpha)}\,\frac{\partial p^{(\beta)}}{\partial y}\ ,\label{A3}
\end{eqnarray}
\begin{eqnarray}
\frac{3}{2}\left (-\frac{\partial a_1}{\partial x}+P_0^{-1}\frac{\partial P_0}{\partial x}a_1+P_0^{-1}\frac{\partial P_0}{\partial y}b_1\right )&=&P_0^2R_{(\alpha)(\beta)(\gamma)(4)}\, \frac{\partial p^{(\alpha)}}{\partial x}\,p^{(\beta)}\,\frac{\partial p^{(\gamma)}}{\partial x}\ ,\label{A4}\\
\frac{3}{2}\left (-\frac{\partial b_1}{\partial y}+P_0^{-1}\frac{\partial P_0}{\partial y}b_1+P_0^{-1}\frac{\partial P_0}{\partial x}a_1\right )&=&P_0^2R_{(\alpha)(\beta)(\gamma)(4)}\,
\frac{\partial p^{(\alpha)}}{\partial y}\,p^{(\beta)}\,\frac{\partial p^{(\gamma)}}{\partial y}\ ,\label{A5}\\
-\frac{\partial b_1}{\partial x}+\frac{1}{2}P_0^{-1}\frac{\partial P_0}{\partial x}b_1-\frac{1}{2}\frac{\partial a_1}{\partial y}-\frac{1}{2}P_0^{-1}\frac{\partial P_0}{\partial y}a_1&=&
P_0^2R_{(\alpha)(4)(\beta)(\gamma)}\,\frac{\partial p^{(\alpha)}}{\partial x}\frac{\partial p^{(\beta)}}{\partial y}\,p^{(\gamma)}\ ,\label{A6}\\
-\frac{\partial a_1}{\partial y}+\frac{1}{2}P_0^{-1}\frac{\partial P_0}{\partial y}a_1-\frac{1}{2}\frac{\partial b_1}{\partial x}-\frac{1}{2}P_0^{-1}\frac{\partial P_0}{\partial x}b_1&=&
P_0^2R_{(\alpha)(4)(\beta)(\gamma)}\,\frac{\partial p^{(\alpha)}}{\partial y}\frac{\partial p^{(\beta)}}{\partial x}\,p^{(\gamma)}\ ,\label{A7}\\
\frac{3}{2}P_0^{-4}b_1+\frac{1}{2}P_0^{-2}\frac{\partial}{\partial x}\left\{P_0^2\left (\frac{\partial}{\partial x}(P_0^{-2}b_1)-\frac{\partial}{\partial y}(P_0^{-2}a_1)\right )\right\}&=&R_{(\alpha)(4)(\beta)(\gamma)}\,\frac{\partial p^{(\alpha)}}{\partial x}\,\frac{\partial p^{(\beta)}}{\partial x}\,\frac{\partial p^{(\gamma)}}{\partial y}\ ,\label{A8}\\
\frac{3}{2}P_0^{-4}a_1+\frac{1}{2}P_0^{-2}\frac{\partial}{\partial y}\left\{P_0^2\left (\frac{\partial}{\partial y}(P_0^{-2}a_1)-\frac{\partial}{\partial x}(P_0^{-2}b_1)\right )\right\}&=&R_{(\alpha)(4)(\beta)(\gamma)}\,\frac{\partial p^{(\alpha)}}{\partial y}\,\frac{\partial p^{(\beta)}}{\partial y}\,\frac{\partial p^{(\gamma)}}{\partial x}\ ,\label{A9}\\
\frac{\partial}{\partial y}(P_0^{-2}\alpha_2)-\frac{\partial}{\partial x}(P_0^{-2}\beta_2)-P_0^{-2}\frac{\partial q_2}{\partial y}&=&\frac{1}{2}R_{(\alpha)(\beta)(\gamma)(\sigma)}\,
\frac{\partial p^{(\alpha)}}{\partial x}\,p^{(\beta)}\,\frac{\partial p^{(\gamma)}}{\partial x}\,\frac{\partial p^{(\sigma)}}{\partial y}\ ,\label{A10}\\
-\frac{\partial}{\partial x}(P_0^{-2}\alpha_2)-\frac{\partial}{\partial y}(P_0^{-2}\beta_2)-P_0^{-2}\frac{\partial q_2}{\partial x}&=&\frac{1}{2}R_{(\alpha)(\beta)(\gamma)(\sigma)}\,
\frac{\partial p^{(\alpha)}}{\partial y}\,p^{(\beta)}\,\frac{\partial p^{(\gamma)}}{\partial y}\,\frac{\partial p^{(\sigma)}}{\partial x}\ ,\label{A11}\\
\frac{\partial}{\partial x}\left\{P_0^2\left (\frac{\partial}{\partial y}(P_0^{-2}\beta_2)+\frac{\partial}{\partial x}(P_0^{-2}\alpha_2)\right )\right\}&+&\frac{\partial}{\partial y}\left\{P_0^2\left (\frac{\partial}{\partial x}(P_0^{-2}\beta_2)-\frac{\partial}{\partial y}(P_0^{-2}\alpha_2)\right )\right\}\nonumber\\
=-P_0^{-2}(\Delta q_2+6\,q_2)&-&P_0^2R_{(\alpha)(\beta)(\gamma)(\sigma)}\,\frac{\partial p^{(\alpha)}}{\partial x}\,\frac{\partial p^{(\beta)}}{\partial y}\,\frac{\partial p^{(\gamma)}}{\partial x}\,\frac{\partial p^{(\sigma)}}{\partial y}\ ,\label{A12}\\
-\frac{1}{2}P_0^{-2}\left\{\frac{\partial}{\partial y}\left (P_0^2\,\frac{\partial c_2}{\partial x}\right )+\frac{\partial}{\partial x}\left (P_0^2\,\frac{\partial c_2}{\partial y}\right )\right\}&=&
R_{(\alpha)(4)(\beta)(4)}\,\frac{\partial p^{(\alpha)}}{\partial x}\,\frac{\partial p^{(\beta)}}{\partial y}\ ,\label{A13}\\
-2\,c_2-P_0^2\frac{\partial^2c_2}{\partial x^2}-P_0\frac{\partial P_0}{\partial x}\frac{\partial c_2}{\partial x}+P_0\frac{\partial P_0}{\partial y}\frac{\partial c_2}{\partial y}&=&
P_0^2R_{(\alpha)(4)(\beta)(4)}\,\frac{\partial p^{(\alpha)}}{\partial x}\,\frac{\partial p^{(\beta)}}{\partial x}\ ,\label{A14}\\
-2\,c_2-P_0^2\frac{\partial^2c_2}{\partial y^2}-P_0\frac{\partial P_0}{\partial y}\frac{\partial c_2}{\partial y}+P_0\frac{\partial P_0}{\partial x}\frac{\partial c_2}{\partial x}&=&
P_0^2R_{(\alpha)(4)(\beta)(4)}\,\frac{\partial p^{(\alpha)}}{\partial y}\,\frac{\partial p^{(\beta)}}{\partial y}\ ,\label{A15}
\end{eqnarray}
with $\Delta q_2=P_0^2\left (\frac{\partial^2q_2}{\partial x^2}+\frac{\partial^2q_2}{\partial y^2}\right ),$ in (\ref{A12}).
\end{widetext}

\bibliographystyle{unsrtnat}
\bibliography{gravcompwaves_bibliography}
\end{document}